\documentclass[aps,prd,reprint,amsmath,amssymb,nofootinbib]{revtex4-2}

\usepackage{slashed}
\usepackage{feynmp}
\usepackage{tikz-feynman}
\tikzfeynmanset{compat=1.1.0}
\usepackage{graphicx}
\usepackage{hyperref}
\usepackage{ulem}
\usepackage{color}
\usepackage{amsmath,amssymb}
\usepackage{subfigure}
\usepackage{siunitx}
\usepackage[utf8]{inputenc}
\graphicspath{{./images/}}
\usepackage{url}

\begin{document}

\newcommand{\spr}[1]{{\color{red}\bf[SP:  {#1}]}}
\newcommand{\che}[1]{{\color{blue}\bf[CE:  {#1}]}}

\title{Dark-sector modifications to Kerr and Reissner-Nordström black hole evaporation}

\author{Christopher Ewasiuk}
\author{Stefano Profumo}
\affiliation{Department of Physics and Santa Cruz Institute for Particle Physics,
University of California, Santa Cruz, California 95064, USA}

\begin{abstract}
We present that large numbers of light hidden-sector degrees of freedom can reverse
the canonical evaporation hierarchy of Kerr--Newman black holes. When the effective
number of uncharged species exceeds a critical value, the mass-loss rate outpaces
the charge-loss rate, producing effective charge growth during evaporation and delaying
neutralization. We investigate the evolution of black holes during the evaporation
process, emphasizing how these quantities evolve relative to one another. Our study
incorporates the effects of greybody factors, near-extremal conditions, and the
introduction of additional particle species beyond the Standard Model. We demonstrate
that the addition of particle degrees of freedom may significantly alter the evaporation
hierarchy, potentially leading to scenarios in which the effective black hole charge
increases during evaporation. Additionally, we examine the impact of Schwinger pair
production and of superradiance on charged, spinning black hole evaporation. These
findings offer new insights into the complex interplay between different black hole
parameters during evaporation and highlight the importance of considering additional
particle species in the process.
\end{abstract}

\maketitle

\section{Introduction}
\label{sec:intro}

Black hole evaporation through Hawking radiation remains central to theoretical physics since its discovery. As black holes emit thermal radiation due to quantum effects near the event horizon, they lose mass and, depending on type, additional conserved quantities including angular momentum and charge \cite{hawking1975particle,page1976massless}. Understanding these dynamics has profound implications for quantum mechanics, thermodynamics, and the information paradox. This study examines the hierarchy by which black holes dissipate mass, charge, and angular momentum.

Page factors \cite{page1976massless,page1977charged} provide a quantitative framework for describing loss rates of energy, angular momentum, and charge during Hawking radiation. These factors have proven crucial for analyzing evaporation dynamics of Schwarzschild, Kerr, and Reissner-Nordstr\"om (RN) black holes within the semi-classical approximation, which assumes a fixed background metric with quantum mechanical evolution of fields and successfully describes evaporation of sufficiently large black holes where back-reaction effects remain negligible \cite{hawking1975particle}.

\subsection{Black Hole Types and Evaporation Mechanisms}

Black holes are characterized by mass ($M$), angular momentum ($J$), and charge ($Q$). Schwarzschild black holes, being non-rotating and uncharged, lose only mass through isotropic thermal radiation \cite{page1976massless}. Kerr black holes introduce rotational complexity: frame-dragging effects produce anisotropic emission, with angular momentum typically lost much faster than mass through higher-spin particle emission, driving a ``spin-down'' toward Schwarzschild-like states \cite{page1976massless,page1976rotating,taylor1998evaporation}. RN black holes, charged but non-rotating, preferentially emit oppositely charged particles, dissipating charge on much shorter timescales than mass when such particles are thermodynamically available (i.e., when their mass is comparable to or below the hole's temperature) \cite{page1977charged,hod2019penrose}.

These evaporation patterns critically depend on the particle spectrum. The existence of degrees of freedom beyond the Standard Model (SM) can substantially modify the dissipation hierarchy described above. Furthermore, greybody factors---which quantify deviations from ideal blackbody spectra due to scattering in curved spacetime \cite{boonserm2012bounding,dong2015gravitational}---play a significant role in determining emission rates and energy partitioning across particle types and modes.

\subsection{Comparative Analysis and Extremal Effects}

While individual black hole types have been studied extensively, systematic comparative analyses using Page factors remain sparse. Schwarzschild black holes evaporate slowest due to isotropic Hawking radiation \cite{page1976massless,masina2021dark}. Kerr black holes lose angular momentum several times faster than mass, with superradiance effects enhancing energy emission from ergoregions \cite{page1976rotating,masina2021dark}. RN black holes rapidly shed charge, subsequently suppressing radiation through Coulomb potential barriers \cite{page1977charged}.

High charge-to-mass ratios ($Q/M$) in RN black holes induce distinctive phenomena including Schwinger pair production and quantum charge fluctuations near event horizons \cite{dong2015gravitational,hod2019penrose}. Near-extremal configurations---whether Kerr or RN black holes approaching physical limits---exhibit vanishing surface gravity, suppressing thermal emission within semi-classical approximations, though quantum corrections modify this behavior \cite{brown2024evaporation,hawking1997extremal}. Additionally, superradiance enables light bosons to form ``gravitational atoms'' around rotating black holes, anomalously accelerating angular momentum loss for specific combinations of spin and boson mass.


\subsection{Motivation and Scope}

Despite detailed studies of individual black hole evaporation characteristics, significant gaps remain in understanding how mass, angular momentum, and charge evolve comparatively across black hole types. Key questions include:
\begin{itemize}
    \item How do Page factors for energy, angular momentum, and charge loss differ quantitatively across Schwarzschild, Kerr, and RN black holes?
    \item What are the relative timescales for losing conserved quantities, and how do greybody factors influence these dynamics?
    \item How do near-extremal conditions (high $a/M$ or $Q/M$) modify evaporation processes?
    \item How do degrees of freedom beyond the Standard Model alter dissipation patterns?
    \item What additional mechanisms can modify relative loss rates of charge, angular momentum, and mass?
\end{itemize}

This study quantitatively addresses these questions. Section~2 describes Page factor evolution for light black holes, outlining computational procedures for time-evolved factors with spin and charge evolution, including effects of dark degrees of freedom. Section~\ref{hierarchy_sect} discusses how relative dissipation of mass, charge, and spin varies with additional particle species. Section~\ref{SchwingerSect} examines time evolution under Schwinger pair production, which significantly impacts light charged black holes. The penultimate section analyzes how superradiance and gravitational atom formation affect angular momentum extraction and Kerr black hole lifetimes. Section~\ref{sec:conclusions} presents final considerations and conclusions. Throughout, we work in natural units where $G = \hbar = c = 1$.

\section{Page Factor Evolution}
The evolution of intrinsic BH parameters such as spin and mass can be determined by the so-called Page factors, as outlined in \cite{Page_rotating}.
Generally, the lifetime of a BH scales as the black hole mass to the third power, $M^3$, given that the evaporation rate follows the classical Stefan-Boltzmann equation, and area is proportional to $R^2_S\sim M^2$, $R_S$ being the black hole Schwarzschild radius, and $T_H\sim M^{-1}$, with $T_H$ the Hawking temperature; as a result, the scale-invariant Page factor for the BH mass is defined as
\begin{equation}
    f(M,x^*) = - M^3 \frac{d \ln(M)}{dt} = -M^2 \frac{dM}{dt} 
\end{equation}
where $x^*$ corresponds to the intrinsic spin $a^*$ or charge $Q^*$, depending on if we are describing the Kerr or RN metrics. We will use here the effective dimensionless quantities $a^* = \frac{a}{M}= \frac{J}{M^2}$ and $Q^* = \frac{Q}{M}$. 

The Page factor for angular momentum is defined as  
\begin{equation}
    g(M,a^*) = - M^3 \frac{d \ln(J)}{dt} = -\frac{M}{a^*} \frac{dJ}{dt} 
\end{equation}
To describe charge evolution, we  define the additional scale-invariant quantity
\begin{equation}
    h(M,Q^*)  = - M^3 \frac{d \ln(Q)}{dt} = -\frac{M^2}{Q^*} \frac{dQ}{dt}. 
    \label{charge_page}
\end{equation}

\begin{figure}[t!]
\begin{center}
\includegraphics[width = 0.4\textwidth, height = 0.4\textwidth ]{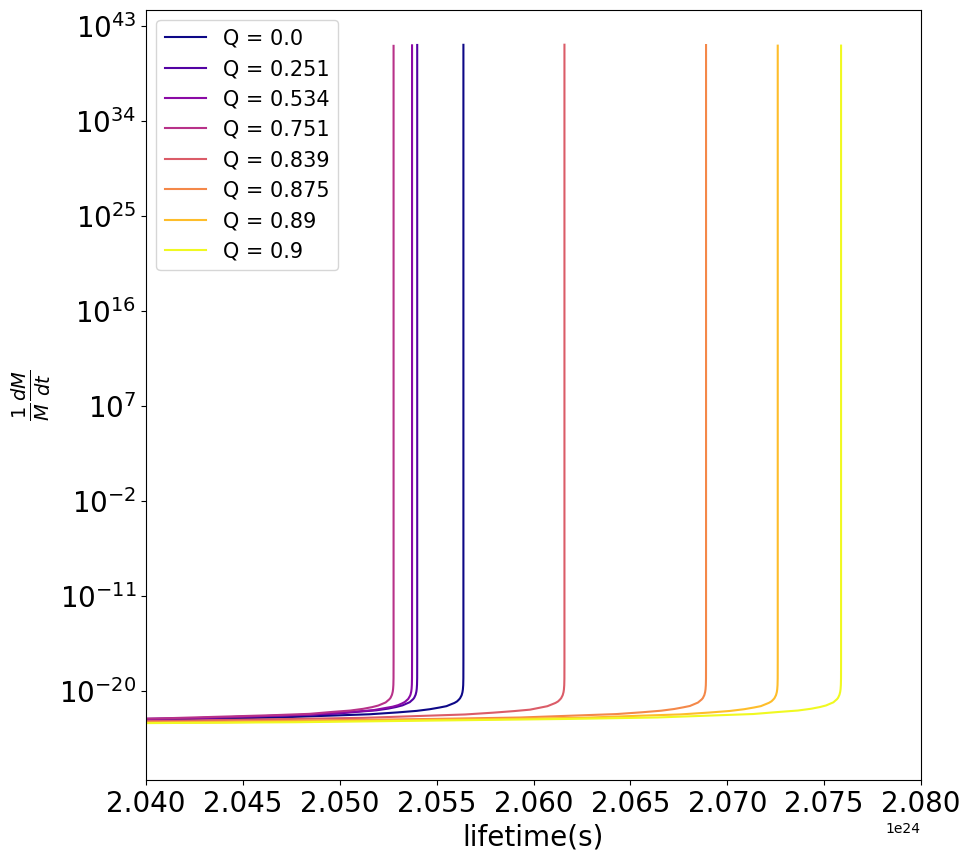}
\caption{Lognormalized mass loss rates as a function of lifetime for different effective charge. Increasing effective charge lowers the overall temperature of the BH, slowing particle emission. The initial BH mass is set at $10^{17}$ g as larger masses will not significantly change the initial mass loss rate.}
\label{fig:dMdt_charge}
\end{center}
\end{figure}

\begin{figure}[t!]
\includegraphics[width = 0.4\textwidth, height = 0.4\textwidth ]{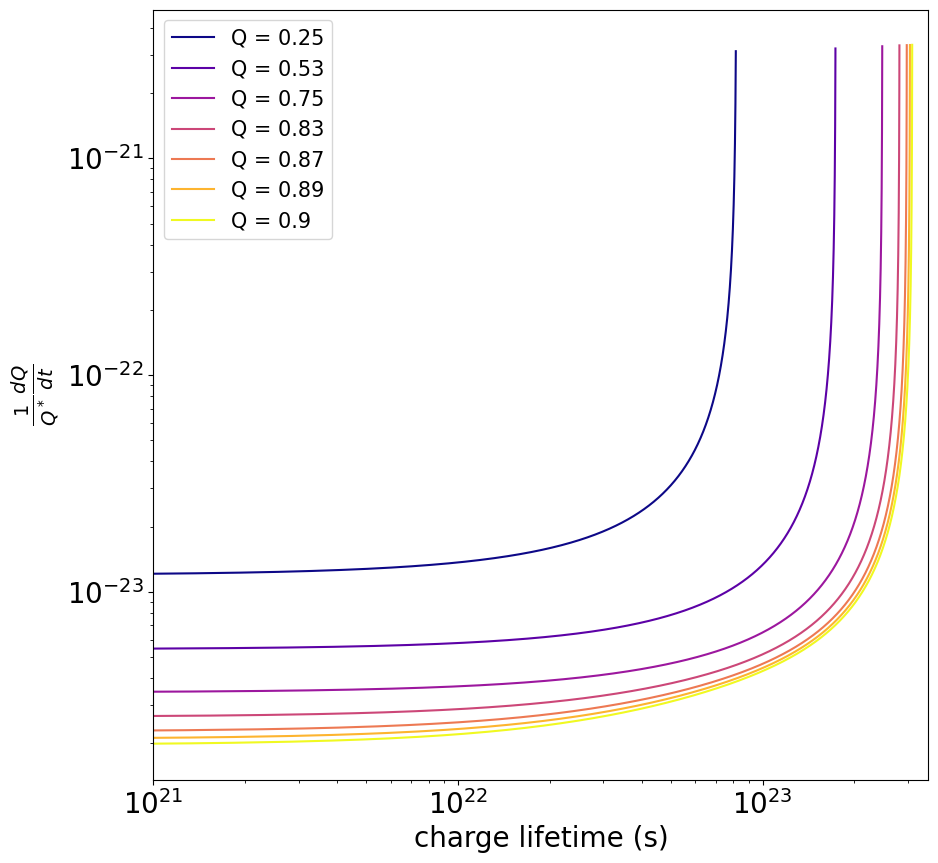}
\caption{Charge loss rate due to Hawking radiation for various spins normalized by the initial effective charge $Q^*$ of a BH.}
\label{fig:dQdt_vs_t}
\end{figure}

\noindent The rates of particle emission are described as in Ref.~\cite{Hawking_BHE} and give

\begin{equation}
    f(M,x^*) = -M^2 \frac{dM}{dt} = -M^2 \int_0^{\infty} E \sum_i\frac{d^2N_i}{dtdE}dE,
    \label{1}
\end{equation}

\begin{equation}
    g(M,a^*) =-\frac{M}{a^*} \frac{dJ}{dt} = -\frac{M}{a^*} \int_0^{\infty}  \sum_i g_i \sum_{l,m}\frac{d^2N_{l,m}}{dtdE}dE, \label{2}
\end{equation}
where the terms 
\begin{equation}
    \frac{d^2N_{l,m}}{dtdE} = \frac{1}{2 \pi} \frac{\Gamma_{s,lm}(M,E,x^*)}{e^{E/T} - (-1)^{2s}},
    \label{N particles}
\end{equation}

\begin{equation}
    \frac{d^2N_{i}}{dtdE} = g_i\sum_{l,m} \frac{d^2N_{l,m}}{dtdE}
\end{equation} describe the total number of outgoing particles created within the BH horizon, as described in \cite{Hawking_BHE}.
This leads us to an analogous Page factor describing charge evolution from Eq.~\eqref{charge_page}:

\begin{equation}
    h(M,Q^*) =-\frac{M^2}{Q^*} \frac{dQ}{dt} = -\frac{M^2}{Q^*} \int_0^{\infty}  \sum_i q_i g_i \sum_{l,m}\frac{d^2N_{l,m}}{dtdE}dE. \label{3}
\end{equation}

\noindent The terms $q_i$ are the individual particle charge and $g_i$ are available particle degrees of freedom. Note that the term $h(M,Q^*)$ does not include a double counting of the number of degrees of freedom from including antiparticles, as antiparticles and particles included would have opposite charge. We define the effective charge as being from 0 to 1, as the process for shedding negative charge is identical to the scenario of shedding positive charge aside from a sign change.

The computations in the following section are done using \textit{BlackHawk} \cite{Blackhawk} for the Kerr metric. However, \textit{BlackHawk} does not account for time evolution of a RN BH and a charged metric adaptation of the code was written to overcome this issue. The analysis code used in this work is openly available online\footnote{\url{https://github.com/cewasiuk/Spin-and-Charge-BH-Evolution}}. 
In our consideration of the charged metric, we have adopted the optical limit approximation to compute the particle's greybody factor: 

\begin{equation}
    \Gamma = 27  M^2 E^2,
\end{equation}
where $E$ is the total energy of the emitted particle,  taken to be extremely large in comparison to an individual particles rest mass.  This approximation becomes increasingly accurate in the high-frequency regime ($\omega M \gg 1$), and is therefore  suited to the extremely light black holes considered in our study, whose high Hawking temperatures push the emission spectrum toward the geometric-optics limit. While not exact, this approach captures the qualitative behavior of RN evaporation in the presence of additional massless degrees of freedom.
The optical-limit treatment leads to modest deviations in the Page factors and produces an overall shortening of the total lifetime relative to the fully tabulated greybody factors. Although our approximation slightly overestimates the evaporation rate, it accurately reproduces the qualitative trends relevant for exploring how large hidden sectors modify RN black hole evolution.

\section{New Results and Physical Implications}

In this work we identify several new features in the standard evaporation hierarchy of black holes—where charge is typically lost first, followed by spin and finally mass—when the evolution is extended to include large numbers of beyond–Standard Model particle species. Our principal results are as follows:

\begin{enumerate}
    \item \textbf{Reversal of charge-loss hierarchy} \\
    For Kerr--Newman black holes coupled to large hidden sectors ($N_{\mathrm{dof}} \gtrsim 100$), the standard evaporation hierarchy is inverted. The mass-loss rate exceeds the charge-loss rate,
    \[
    \left|\frac{\dot{M}}{M}\right| > \left|\frac{\dot{Q}^*}{Q^*}\right|,
    \]
    leading to effective charge growth during evaporation. This contrasts with previous expectations that charged black holes always neutralize first.

    \item \textbf{Critical threshold for reversal} \\
    We derive a quantitative condition for this inversion, showing that it occurs when
    \[
    N_{\mathrm{dof}} > N_{\mathrm{crit}} \approx 220,
    \]
    for beyond Standard model massless degrees of freedom, beyond which energy loss to uncharged species dominates over electromagnetic discharge.

    \item \textbf{Delayed or halted neutralization} \\
    In the high-$N_{\mathrm{dof}}$ regime, black holes can approach near-extremal states ($Q^*/M \to 1$) without rapid charge loss. These configurations exhibit prolonged lifetimes due to suppressed Hawking temperatures and reduced Schwinger emission.

    \item \textbf{Interplay with rotation} \\
    For Kerr configurations, spin-down remains dominant, but frame-dragging modifies the effective emission balance, coupling angular momentum loss to charge evolution. This coupling introduces a correlated decay track in the $(a^*, Q^*)$ phase space not captured in prior treatments.

    \item \textbf{Reemergence of discharge via Schwinger pair production} \\
    In the case of a large dark sector or light extremal black holes ($M \lesssim 10^{17}\,\mathrm{g}$), quantum electrodynamic discharge again dominates, restoring neutrality and terminating the reversed regime.
    
    \item \textbf{Superradiant extraction of angular momentum} \\
    We incorporate the effects of black hole superradiance into the evaporation hierarchy, showing that massive bosonic fields can remove angular momentum far more efficiently than Hawking radiation. 

    \item \textbf{Numerical confirmation} \\
    Numerical solutions to the coupled Page-factor equations confirm these behaviors, delineating regions in $(M, Q^*, a^*)$ space where charge growth occurs and illustrating the modified evaporation trajectories under large hidden-sector particle counts.
\end{enumerate}

Together these results demonstrate that the canonical evaporation sequence, charge loss preceding mass loss, is not inherently universal but depends sensitively on the particle content of the theory. Large hidden sectors can qualitatively change black hole evolution, producing transiently charged or near-extremal relics with potential implications for early-universe black hole populations and dark-sector phenomenology.

\begin{figure}[t!]
\includegraphics[width = 0.4\textwidth, height = 0.4\textwidth ]{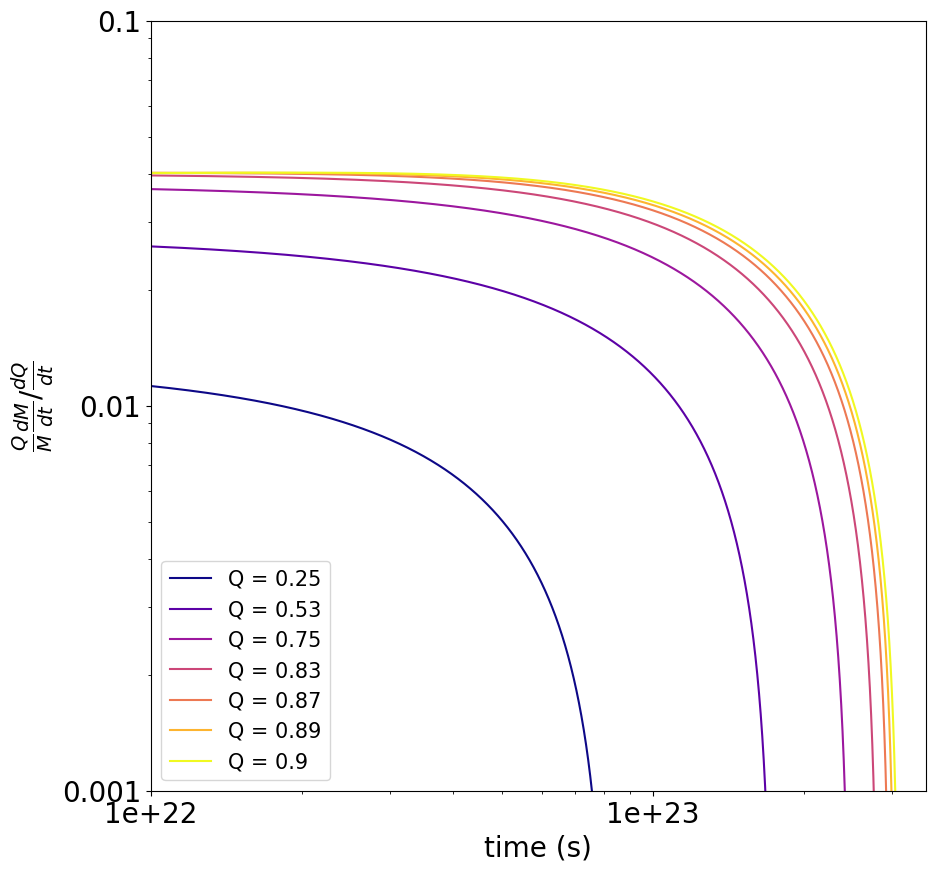}
\caption{ Lines of $\frac{\frac{1}{M} dM/dt}{\frac{1}{Q^*} dQ^*/dt} $ lognormalized for a RN BH for various effective charge values. Initial conditions are for a $M = 10^{17} g$ BH and no included DM. }
\label{fig:lognorm_dMdt_dQdt}
\end{figure}

\section{Effect of a large dark sector on parameter evaporation}

Introducing a dark sector containing a large number of degrees of freedom results in significant  changes in the evolution of charged or rotating BHs. The effective spin and charge of a BH are defined as $a^* = \frac{J}{M^2}$ and  $Q^* = \frac{Q}{M}$, where an effective spin or charge equal to 1 corresponds to extremality where evaporation shuts off.  The total number of particles needed to neutralize a charged black hole are of order $M$, whereas the number needed to carry off overall angular momenta $J$ scales as  $M^2$. Therefore the standard hierarchy of parameter loss, when considering purely evaporative processes, begins with charge neutralization, followed by loss of spin and finally mass evaporation which, as noted above, scales as $M^{-2}$. Rather than comparing baseline cases, we target regimes that perturb the hierarchy itself. Starting with charged black holes, we include a substantial population of uncharged dark-sector species and quantify the parameter ranges where $f(M,Q^*)$ surpasses $h(M,Q^*)$, signaling hierarchy reversal.

\subsection{Reissner-Nordström Evaporation Rates}

The overall effective charge of a black hole (BH) shrinks both the horizon and Cauchy radii according to
\begin{equation}
    r_{\pm} = r_{H,C} = r_S \frac{1 \pm \sqrt{1 - {Q^*}^2}}{2},
    \label{radii}
\end{equation}
where $r_S = 2M$. This modifies the black hole temperature,
\begin{equation}
    T_{RN} =  \frac{r_+ - r_-}{4 \pi r_+^2},
    \label{RNTemp}
\end{equation}
and thus alters the emission spectrum. As $T_{RN}$ increases, the probability of emitting particles of a given rest mass rises, as reflected in Eq.~\eqref{N particles}. 

Figure~\ref{fig:dMdt_charge} shows that black holes with higher effective charge have significantly longer lifetimes for the same mass. This occurs because a larger charge reduces the temperature, thereby suppressing particle emission.

At the same time, stronger charge enhances the absolute rate of charge loss. When this rate is normalized by the effective charge, however, we find that black holes with smaller initial $Q^*$ lose charge more efficiently. This behavior, illustrated in Figs.~\ref{fig:dQdt_vs_t} and~\ref{fig:lognorm_dMdt_dQdt}, explains why highly charged black holes require longer timescales to neutralize. For sufficiently light black holes capable of thermally emitting electrons, the charge-loss rate can exceed the mass-loss rate, indicating that charge depletion dominates the early evaporation phase.

\begin{figure*}[t!]
    \centering
    \includegraphics[width=0.9\linewidth]{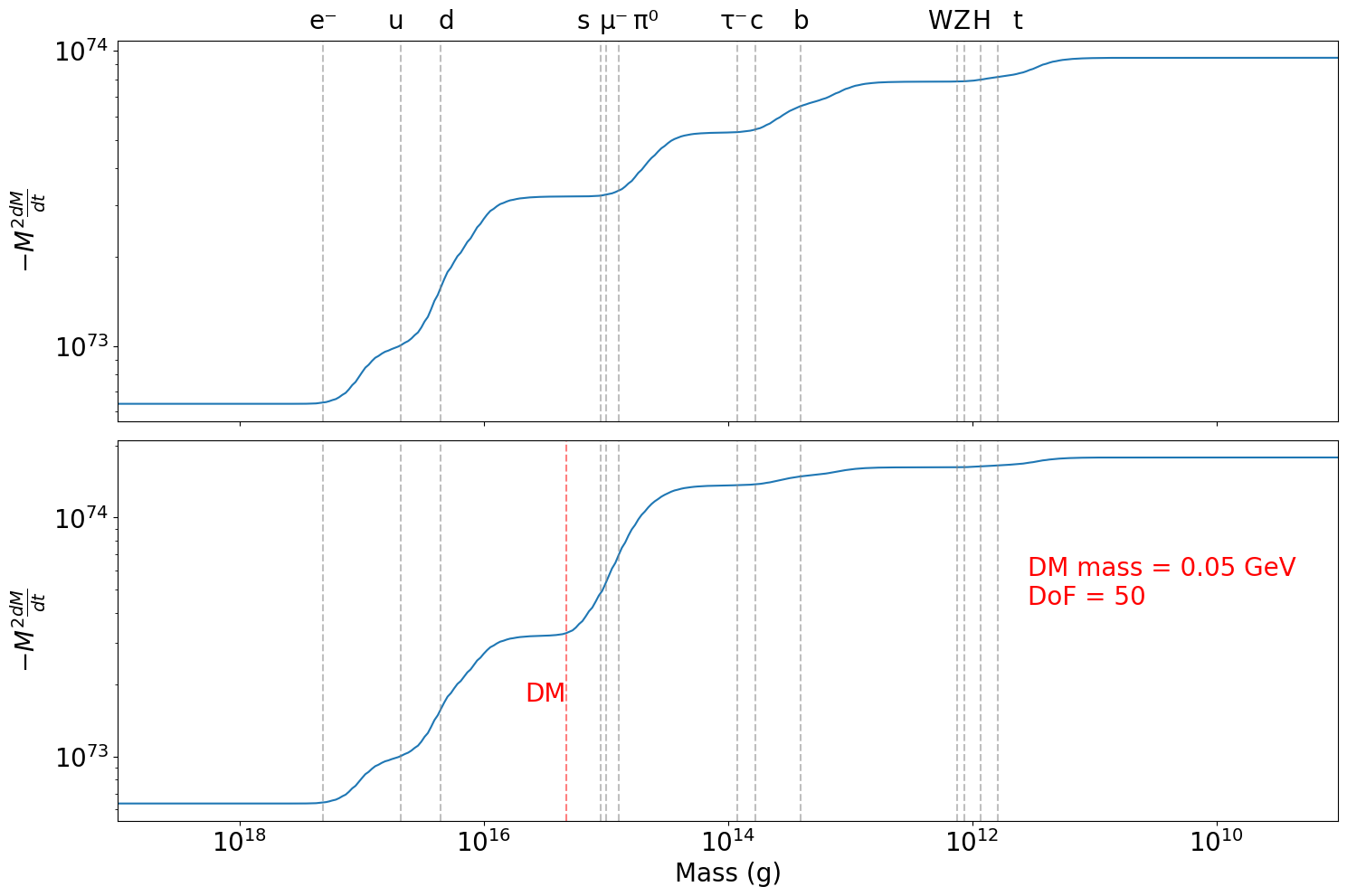}
    \caption{Approximate black hole mass threshold for massive particles to significantly contribute to the normalized mass loss rate. All massive particles have a nonzero emission rate at all times, but are exponentially supressed by their respective greybody factor. Significant emission occurs when $T_{BH} \approx m_{p}$. Influence of a light, neutral dark sector on the normalized mass loss rate is included to illustrate its effect.}
    \label{fig:Hawking_Threshold}
\end{figure*}

To compare the relative rates of mass and charge loss, we define the log-normalized ratio (LNR)
\begin{equation}
\frac{\frac{1}{M} \frac{dM}{dt}}{\frac{1}{Q^*} \frac{dQ^*}{dt}},
\label{LNR}
\end{equation}
which tracks the time evolution of both quantities on a dimensionless scale. The effective charge derivative is given by
\begin{equation}
    \frac{dQ^*}{dt}  =  \frac{1}{M} \frac{dQ}{dt} - \frac{Q^*}{M} \frac{dM}{dt}, 
    \label{crossover_eq}
\end{equation}
taking care to distinguish between the true charge-loss rate $\frac{dQ}{dt}$ and the effective charge-loss rate $\frac{dQ^*}{dt}$. Written in terms of the Page factors $f(M,Q^*)$ and $h(M,Q^*)$ defined in Eqs.~\eqref{1} and~\eqref{3}, Eq \eqref{crossover_eq} becomes
\begin{equation}
\frac{dQ^*}{dt}  = Q^* \frac{f(M,Q^*) - h(M,Q^*)}{M^3}.
\label{effective_charge_loss}
\end{equation}

Our analysis focuses on the relative scaling between the mass and charge-loss rates. Both Page factors depend on the number of available degrees of freedom of the emitted species. However, the Page factor associated with charge, $h(M,Q^*)$, only receives contributions from charged particles. As a result, the emission of neutral species enhances the mass and spin Page factors while leaving $h(M,Q^*)$ unchanged.

\begin{figure}
\begin{center}
\includegraphics[width = 0.4\textwidth, height = 0.4\textwidth ]{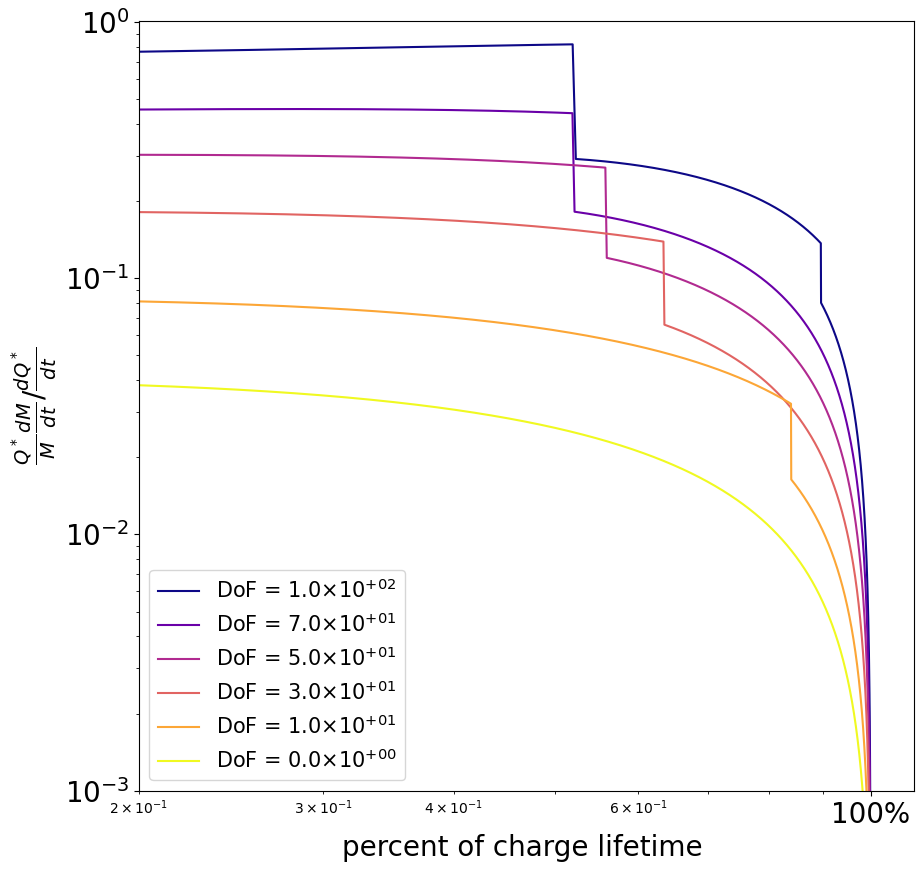}
    \caption{ Illustration of the effects of adding spin 0, chargeless DM to the BH evaporation spectrum while varying the number of degrees of freedom. Initial conditions are for a BH with mass $1^{17} g$, effective charge $Q^* = 0.9$ and introduced dark matter with $m_{DM} = 0 $ GeV. Adding larger degrees of freedom will result in a transition from a regime where $\frac{1}{Q}\frac{dQ}{dt}$ is dominant to a regime where$\frac{1}{M}\frac{dM}{dt}$ begins to dominate. Such a transition is illustrated for the same initial parameters in Fig. \ref{fig:crossover}, and occurs when the lines reach a maximum of 1. Illustration only shows rates when the BH is considered charged, afterwards the BH is considered static.}

\label{fig:lognorm_dMdt_dQdt_DOF}
\end{center}

\end{figure}

In order to invert the typical evaporation sequence in which charge neutralization precedes mass loss, we introduce massless, neutral degrees of freedom to create a regime in which the black hole loses mass faster than charge. The introduction of massless particles is chosen so that the mass loss rate can be enhanced prior to the emission of any charged particles, such as the electron. The black hole mass is taken to be relatively light, at $10^{17}$ g—approximately the scale at which the emission of massive particles becomes significant. The corresponding mass thresholds at which all Standard Model species are emitted at non-negligible rates is included for reference in Fig.~\ref{fig:Hawking_Threshold}. We find that if the number of new degrees of freedom is less than $N \approx 200$, $\frac{dQ^*}{dt}$ remains negative and the black hole continues to evaporate normally without any effective charge growth. The impact of introducing dark sectors of this scale are illustrated in Fig.~\ref{fig:lognorm_dMdt_dQdt_DOF}. At early times, the true charge-loss rate $\frac{dQ}{dt}$ stays nearly constant (extremely small, but nonzero due to a nonzero emission probability for any masssive particle), while the mass-loss rate $\frac{dM}{dt}$ increases as new dark-sector species contribute to evaporation. As the black hole evolves, charged particle emission becomes more frequent, enhancing $\frac{dQ}{dt}$ and producing the downward trend in the log-normalized ratio. For the modest number of additional degrees of freedom shown in Fig.~\ref{fig:lognorm_dMdt_dQdt_DOF}, mass loss continues to dominate over charge loss throughout the evolution.

For the parameters considered, the LNR (Eq. \eqref{LNR} approaches unity when the normalized mass and effective charge loss rates become comparable. In our simulation this occurs at the upper limit of the degrees of freedom considered above, or $N =200$. Around this value charge loss and mass loss proceed at similar rates, leading the effective charge $Q^*$ to temporarily plateau. 

\begin{figure}[t!]
\begin{center}
\includegraphics[width = 0.4\textwidth, height = 0.4\textwidth ]{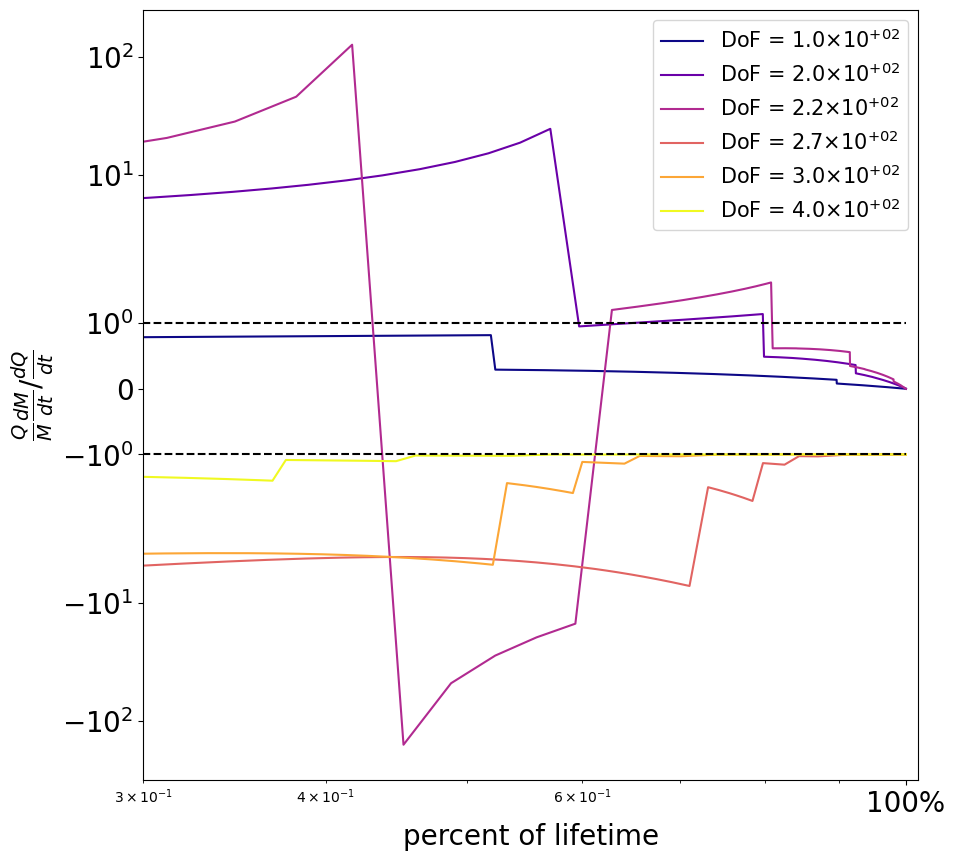}
\caption{Lognormalized ratio of mass and charge loss for chargeless degrees of freedom starting when mass loss dominates and transitioning to a region where effective charge loss dominates. Effective charge loss rates close to zero result in the numerical divergences, and are moments when charge loss rates match mass loss rates.}
\label{fig:crossover}
\end{center}
\end{figure}

Degrees of freedom higher $N \simeq 200$ always result in some duration of a BH's lifetime where $\frac{dM}{dt}$ is dominant in comparison to $\frac{dQ}{dt}$. When this number of additional degrees of freedom are introduced, some portion of the BH's lifetime will be spent increasing effective charge. Additionally, we also find that the entirety of a BH's lifetime will be spent with effective charge increasing for degrees of freedom greater than $N \simeq 270$, which would result in a maximally charged BH with a temperature that approaches zero. Using Eq \eqref{crossover_eq}. We can verify this by looking at the limit as $N \rightarrow \infty$ for Eq. \eqref{crossover_eq}
\begin{equation}
\lim_{N \rightarrow \infty}\big(\frac{1}{M} \frac{dQ}{dt} - \frac{Q^*}{M} \frac{dM}{dt}\big)\approx  - \frac{Q^*}{M} \frac{dM}{dt}, 
\end{equation}

since increasing the number of massless, neutral degrees of freedom will only increase $\frac{dM}{dt}$. The LNR will then reduce to :
\begin{equation}
   \frac{\frac{1}{M} \frac{dM}{dt}}{\frac{1}{Q^*} \frac{dQ^*}{dt}} = \frac{\frac{1}{M} \frac{dM}{dt}}{ \frac{1}{Q^*} \big( -\frac{Q^*}{M} \frac{dM}{dt} \big)} = -1  \label{massdom}
\end{equation}
for large values of $\frac{dM}{dt}$. This effect can be seen within Fig. \ref{fig:crossover}, where the values for $\frac{dQ^*}{dt}$ transition from negative to positive.  This causes the effective charge of the BH to increase after the introduction of a large number of new, non-charged degrees of freedom. For introduced degrees of freedom that allow $f(M,Q^*)$ to become comparable to $h(M,Q^*)$, the effective charge neutralization of the BH stagnates and both true charge and mass evaporate at comparable rates. 

For a BH of mass $10^{17}$g, we estimate that number of massless degrees of freedom required hovers between roughly $220 \leq N_{dof} \leq 270$ and can be seen in Fig. \ref{fig:crossover}.  The number of degrees of freedom required for this transition is BH mass dependent,  again seen through Eq. \eqref{crossover_eq}. Any introduced number of degrees of freedom larger than this, for a given BH mass, will result in a scenario where the effective charge of the BH will increase. This upsets the traditional hierarchy where charge will be evaporated away first. Varying the mass of the included dark matter serves to increase the BH lifetime; i.e. as the mass of the dark matter increases, the temperature threshold for dark matter evaporation increases, resulting in a longer duration spent radiating non-DM particles.

In this setup we have considered relatively light BHs that can thermally emit massive particles. For smaller masses, the Hawking temperature rises rapidly, allowing emission of progressively heavier species and broadening the emission spectrum. Conversely, heavier black holes have much lower temperatures, limiting emission to the lightest particles—photons and neutrinos—while heavier ones are exponentially suppressed. Most of the lifetime of such massive black holes is therefore spent radiating massless degrees of freedom, which reduces mass but not charge. In this regime, the effective charge $Q^* = Q/M$ increases steadily, always driving the black hole toward extremality \textit{without} the need for an introduced dark sector. This effect is also expected to increase the lifetime of the BH indefinitely. The increase of effective charge lowers the BH temperature, suppressing the mass loss page factor $f(M,Q^*)$. The end result is effectively an extremal black hole with a suppressed mass and charge loss and a Hawking temperature that approaches zero that  any evaporative processes. We will see later that the tendency for a BH to approach more extremal can be negated in the presence of an extreme electric field when we include the effects of Schwinger pair production. The inclusion of this effect in addition to Hawking radiation is talked about in more detail within Section \ref{SchwingerSect}.

\subsection{Kerr Evaporation Rates}

We consider here BHs with intrinsic angular momentum $J$ instead. The greybody factors used in this section were numerically evaluated for each configuration of the Kerr metric, providing more accurate predictions for spin evolution and black-hole lifetimes compared to the charged case.  The parameter describing the evolution of a BH with respect to its angular momentum $J$ is the effective spin parameter $a^*$. Similarly to Eq. \eqref{effective_charge_loss}, we can find the effective spin loss rate 
\begin{equation}
    \frac{da^*}{dt} = \frac{1}{M^2}\frac{dJ}{dt} - 2\frac{J}{M^3}\frac{dM}{dt}
    \label{dadt}
\end{equation}
In terms of Page factors defined in Eqs. \eqref{1}, \eqref{2}, this is
\begin{equation}
\frac{da^*}{dt} = a^* \frac{2f(M,a^*) - g(M,a^*)}{M^3}
\end{equation}
which will dictate the evaporation of spin for a Kerr BH. 

\begin{figure}[t!]
\includegraphics[width = 0.4\textwidth, height = 0.4\textwidth ]{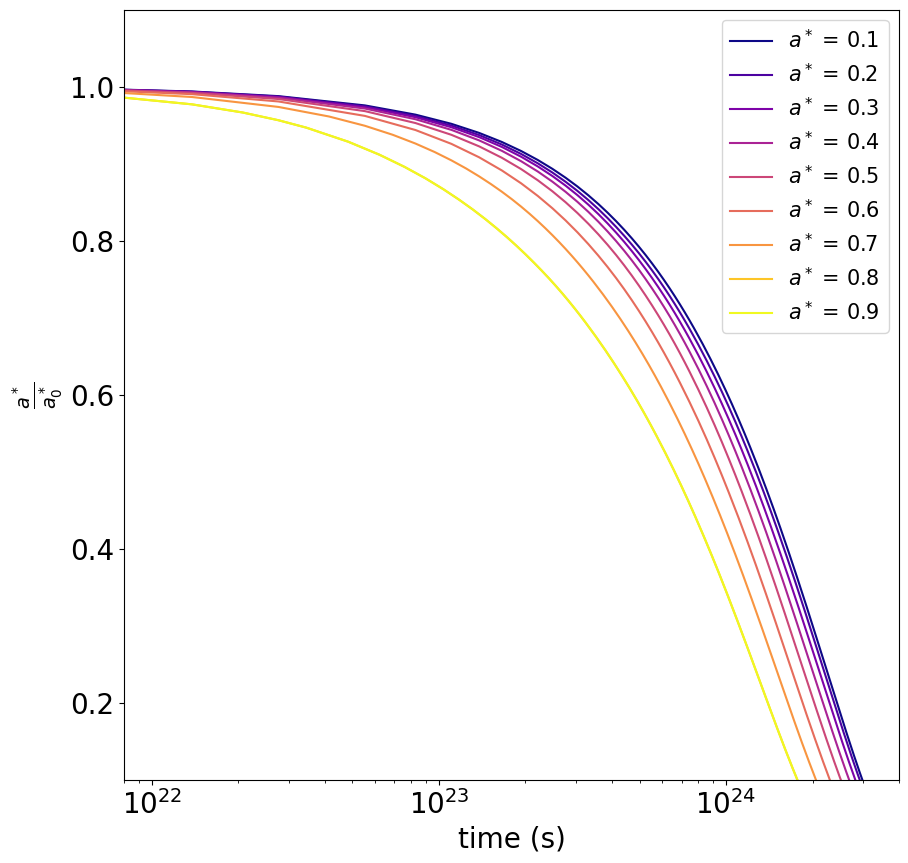}
\caption{Overlay of varying effective spin normalized by initial spin value as a function time for a Kerr BH. Lifetime is given for a Kerr BH of mass $!e17g$, and introduced DM is not considered.}
\label{fig:kerr_spin_v_t}
\end{figure}

Contrary to the case of shedding charge, there is no scenario in which an emitted particle always carries zero angular momentum away from the black hole. In the case of charge loss, the black hole can, in principle, emit a neutral particle, meaning that not every emission necessarily contributes to charge depletion. However, for angular momentum, any emitted particle can carry away some amount of total angular momentum, ensuring that the black hole’s angular momentum is always being affected by its radiation. Even if we consider a particle with a quantum spin number of zero (such as a scalar boson), the total angular momentum $J$ of the emitted particle is given by the sum of its intrinsic spin $S$ and its orbital angular momentum $L$,
\begin{equation}
    J = L +S
\end{equation}
For a spinless particle, $S = 0$, but the emitted particle will still have some nonzero orbital angular momentum $L$. Since angular momentum is quantized, even the lowest possible allowed orbital angular momentum state for a given emission process is nonzero, particularly when considering emissions from a rotating body. This results in a scenario where $g(M,a^*)$ will always be greater than $2f(M,a^*)$ for all greybody factors tabulated. Therefore, spin will always be shed before mass and will cause a BH to tend away from an extremal BH for all particles considered. Still, we can compare timescales of effective spin loss against the charge loss values obtained above. Looking at Fig. \ref{fig:kerr_spin_v_t}, we can see that the timescale for spin evaporation for a maximally spinning ($a^*=0.9$), $10^{17}$ g mass BH is $\approx 1.8 \times 10^{24}$  seconds. Compared to the charge lifetime of $\approx 3 \times 10^{23}$ seconds, we find that spin is shed approximately 6 times slower in our analysis. Again, it is important to emphasize that our lifetime results for charge are an overestimation, however, the hierarchy coincides with what is to be expected. 


\begin{figure*}[t!]
\includegraphics[width = 1.0\textwidth, height = 0.25\textwidth ]{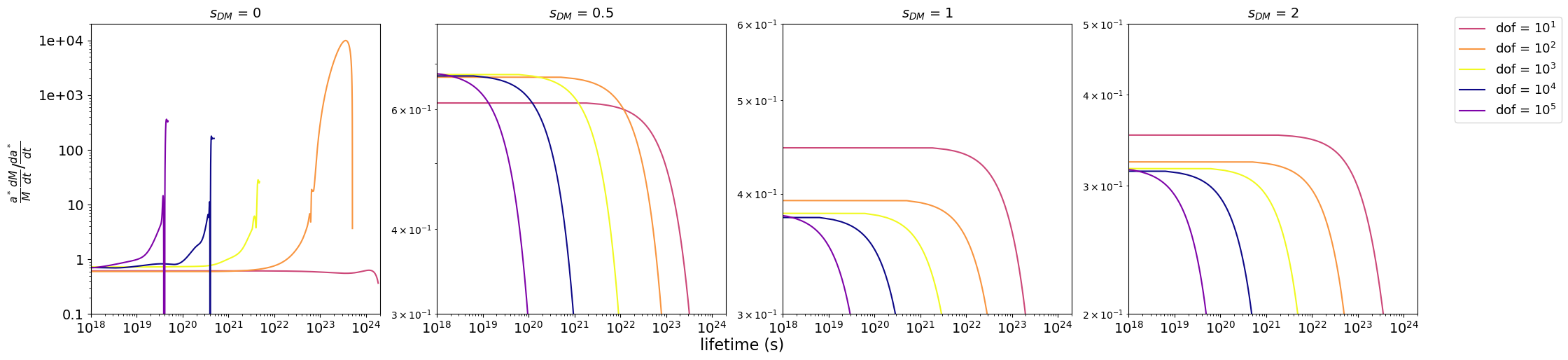}
\caption{Effect of adding various spin dark matter with various masses to the evaporation model of a Kerr BH. BH initial parameters are $a^* = 0.8$, $m_{BH} = 10^{17}g$.  }
\label{fig:spin0_mdm}
\end{figure*}

\begin{figure}[h!]
\includegraphics[width = 0.4\textwidth, height = 0.4\textwidth ]{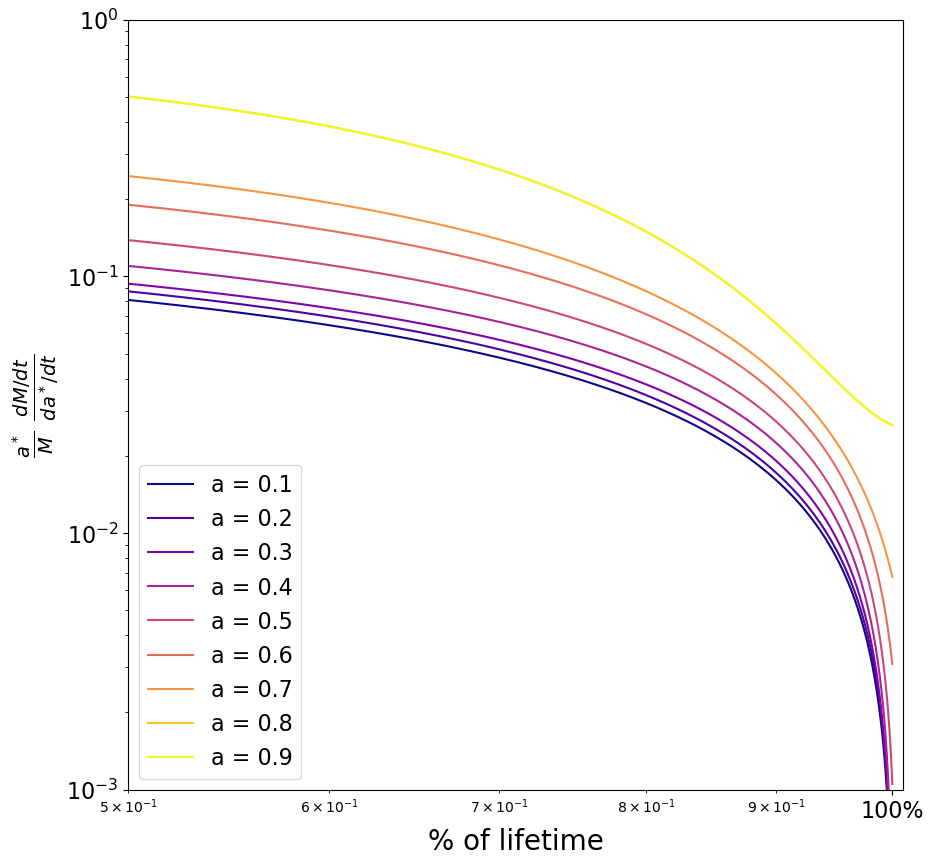}
\caption{Lines of $\frac{\frac{1}{M} dM/dt}{\frac{1}{a^*} da^*/dt} $ lognormalized for a Kerr BH for various effective spin values.}
\label{fig:dadt_vs_t}
\end{figure}

We have also investigated the effects of introducing massless particles of varying degrees of freedom and spin on shedding angular momenta. The results can be seen in Fig. \ref{fig:spin0_mdm}. The most interesting case is the far left panel, where we have considered spin zero particles. In this case, the overall spin carried away by the particle is entirely orbital angular momentum. While the page factor for spin $g(M,a^*)$ is always larger than the mass loss term $2f(M,a^*)$, in this case the difference between the two becomes very small. This results in the denominator term $\frac{da^*}{dt}$ in the lognormalized ratio to become extremely small in comparison to the mass loss term $\frac{dM}{dt}$. For all other cases the intrinsic spin $S$ of the particle acts to increase the spin loss term, resulting in a lognormalized ratio less than 1. We observe apparent divergences in the spin-zero case near the end of the black hole’s spin-down phase. This behavior arises because a scalar field with a large number of degrees of freedom removes angular momentum increasingly inefficiently as the black hole approaches low spin. As $a^*$ decreases, the effective spin-loss rate $\frac{da^*}{dt}$ falls sharply, since there is progressively less angular momentum per emitted quantum. When this declining spin-loss rate is combined with the continually increasing mass-loss rate $\frac{dM}{dt}$, the resulting ratio produces a sharp spike for the $S=0$ case. We do not see this effect for the other cases as having some nonzero spin number $S$ causes the effective spin loss rate to be much more stable even as the BH has a small amount of angular momentum left to radiate.

We have also illustrated how effective spin affects the lognormalized ratio $\frac{a^*}{M} \frac{dM/dt}{da^*/dt}$ in the absence of a large number of introduced degrees of freedom in Fig. \ref{fig:dadt_vs_t}. Again the lognormalized ratio expressed in the figure shows how the normalized mass loss and spin loss rates compare. Values greater than 1 illustrate regions where mass is shed faster than spin, and values less than 1 represent the inverse. In Fig. \ref{fig:dadt_vs_t}, the plot describes how as effective spin of a BH is increased the spin loss rate is decreased in comparison to the mass loss rate. Again this is a result of the dependence of a Kerr BHs temperature on the effective spin of the BH. As effective spin increases the overall BH temperature decreases, resulting in delayed particle evaporation thresholds and a higher exponential suppression of evaporated particles in comparison to lower effective spins. As we approach the end of a BHs lifetime the spin loss term begins to dominate until spin is completely shed.

\section{Relative Hierarchies} \label{hierarchy_sect}

The purpose of the section above is to illustrate the possibility of upsetting the hierarchy of charge, spin and mass loss as dictated by the standard model. Our findings indicate that under any circumstances, spin is lost before mass. However, we have found that through the introduction of a large number of  massive or massless degrees of freedom we can create a scenario where we can increase the effective charge of a RN BH. The standard hierarchy we set out to upset is as follows; charge is neutralized first, followed by spin and finally mass will be depleted ( $ \frac{dM}{dt} < \frac{dJ}{dt} < \frac{dQ}{dt} $). There are two distinct ways in which we can upset this natural order.

The first possibility of changing the order of parameters lost is through the introduction of a select number of chargeless degrees of freedom. In our consideration of a BH of mass $M = 10^{17}$g, if we include massless scalar degrees of freedom within the range of $100 \lessapprox N_{dof} \lessapprox 220$, we find  a scenario where the hierarchy changes and spin is shed first, followed by charge neutralization and then finally mass loss ( $ \frac{dM}{dt} < \frac{dQ}{dt} < \frac{dJ}{dt} $): Since the introduction of  chargeless degrees of freedom has no effect on the rate of charge loss, this introduction solely effects the rates of mass and charge loss. Degrees of freedom larger than $\approx 220 $ result in increasing effective charge, and would make mass evaporate faster than charge. However, degrees of freedom greater than $\approx 100$ will result in spin being lost faster than charge being lost. Again this is a direct result of these introduced degrees of freedom always being able to carry away some nonzero amount of angular momentum. Again, the upper limit of the number of degrees of freedom is set by the BH mass and can be seen in the dependence of $\frac{dQ^*}{dt}$ on $M$ in Eq. \eqref{crossover_eq}. The lower limit is also dependent on the intrinsic spin $S$ of the introduced particle. Higher particle spins will require lower degrees of freedom due to the dependence of the intrinsic spin $S$ in the total angular momentum $J$ carried away from the BH. Lower limits can be extracted by looking at degrees of freedom in Fig \ref{fig:spin0_mdm} the result in Kerr lifetimes lower than $\approx 3 \times 10^{23}$ seconds for specific intrinsic spins. Again as a BH gets heavier the emission spectrum gets exponentially suppressed. So the upper limit will change for select BH masses as well. 

The second possibility is when we consider degrees of freedom higher than $ \approx 220$. In this scenario, we find that the effective charge increases for a $M = 10^{17}$g BH. While the overall charge of the BH goes down, it is outpaced by mass. In this scenario we find that again spin is lost first, but we have the order of mass and charge loss reversed in comparison to the first scenario ( $ \frac{dQ}{dt} < \frac{dM}{dt} < \frac{dJ}{dt} $). 

Note, again, that in the absence of Schwinger pair production extremely massive BHs act similar to blackbodies. In this case, even without the addition of any dark degrees of freedom, we have a scenario where the BH evaporates a significant portion of mass before having any noticeable amount of charge evaporated. Since any charged particle within the standard model has some nonzero mass, the production of such particles will be exponentially suppressed for the majority of the BHs lifetime. Once the BH reaches a mass of roughly $10^{17}$g, it begins to emit a significant amount of charged particles. before that point we find that the effective charge will increase and approach a extremal charged BH. For the majority of the lifetime of a BH, effective charge increases until charged particles can be produced. While this isn't necessarily upsetting the natural hierarchy discussed above since charge is eventually dissipated before mass near the end of the BHs lifetime, the majority of the lifetime is naturally spent increasing the effective charge for heavier BHs.

\section{A Charged Caveat: Schwinger Pair Production }\label{SchwingerSect}
The above analysis for a RN BH only considers the production of charged particles through the computation of their greybody factors outlined in \cite{Blackhawk}. We did not consider, however,  purely classical processes that could neutralize the BH. For example, if the BH were surrounded by charged interstellar media, we would see a fairly quick neutralization of the charge due to selective accretion of particles with charge opposite the BH. Still, even in the absence of such media we can still have an enhanced charge neutralization through quantum effects. A possible significant contribution comes from the Schwinger effect \cite{Schwinger_seminal,Lin:2024aa,Chen:2012zn}. The Schwinger effect describes spontaneous pair creation of charged particles in the presence of an extremely strong electric field. In the majority of electric fields the production is exponentially suppressed, yet for electric fields higher than the threshold field this effect can become dominant. Such a strong field results in vacuum decay and the subsequent creation of electron-positron pairs, serving to lower the charge of the field by $2e$ for each pair created \cite{Schwinger_seminal} \cite{Hiscock:1990ex}. The greybody factors for electron and positron production can then be significantly enhanced via Schwinger pair production.
 
 We can estimate the effect that this has on the charge lifetime of a RN BH by modifying Eq. \eqref{crossover_eq}. Starting with the Schwinger pair production formula \cite{Hiscock:1990ex}, the electron-positron creation rate per unit volume is 
\begin{equation}
    \Gamma = \frac{e^2}{4\pi^3}\,\frac{Q^2}{r_+^4}
    \exp\left(-\frac{\pi m_e^2 r_+^2}{eQ}\right)
    \left[\,1 + \mathcal{O}\!\left(\frac{e^3 Q}{m_e^2 r_+^2}\right) + \cdots \right]
    \label{schwinger_gamma}
\end{equation}
where $m_e$ is the mass of the electron and $Q$ is the BH charge defined as $Q = Q^* M$. We will require that the term $\frac{e^3Q}{m^2r^2}<<1$ and consider only the leading order term. If this assumption is violated the series is no longer convergent and can be assumed to approach infinity, resulting in a massive discharge of the RN BH until the effective charge is low enough to allow the condition to be satisfied. The charge loss rate per unit time can be obtained via integration 
\begin{equation}
    \frac{dQ_S}{dt} = -2e \int \Gamma dV,
\end{equation}

where $Q_S$ has been introduced to denote the difference between charge loss specifically due to Schwinger pair production. Integrating Eq. \ref{schwinger_gamma} rate per unit volume we find the charge loss rate to be
\begin{equation}
    \frac{dQ_S}{dt} = \frac{-e^3}{\pi^2} \frac{1}{r_+} \exp{\bigg[ -\frac{r_+^2}{QQ_0}} \bigg] - \frac{\pi}{(QQ_0)^{1/2}} \text{erfc}\bigg[ \frac{r_+}{(QQ_0)^{1/2}}\bigg] .
\end{equation}
where $\text{erfc}(x)$ is the complementary error function and $Q_0 = \frac{ e}{\pi m_e^2}$. Combining the effects of Schwinger pair production, the effective charge loss in Eq \eqref{crossover_eq} becomes 
\begin{equation}
        \frac{dQ*}{dt}  =  \frac{1}{M} \frac{dQ}{dt} +  \frac{1}{M} \frac{dQ_S}{dt} - \frac{Q^*}{M} \frac{dM}{dt} \label{Schwinger_dQdt}
\end{equation}

\begin{figure}[t!]
\includegraphics[width = 0.4\textwidth, height = 0.4\textwidth ]{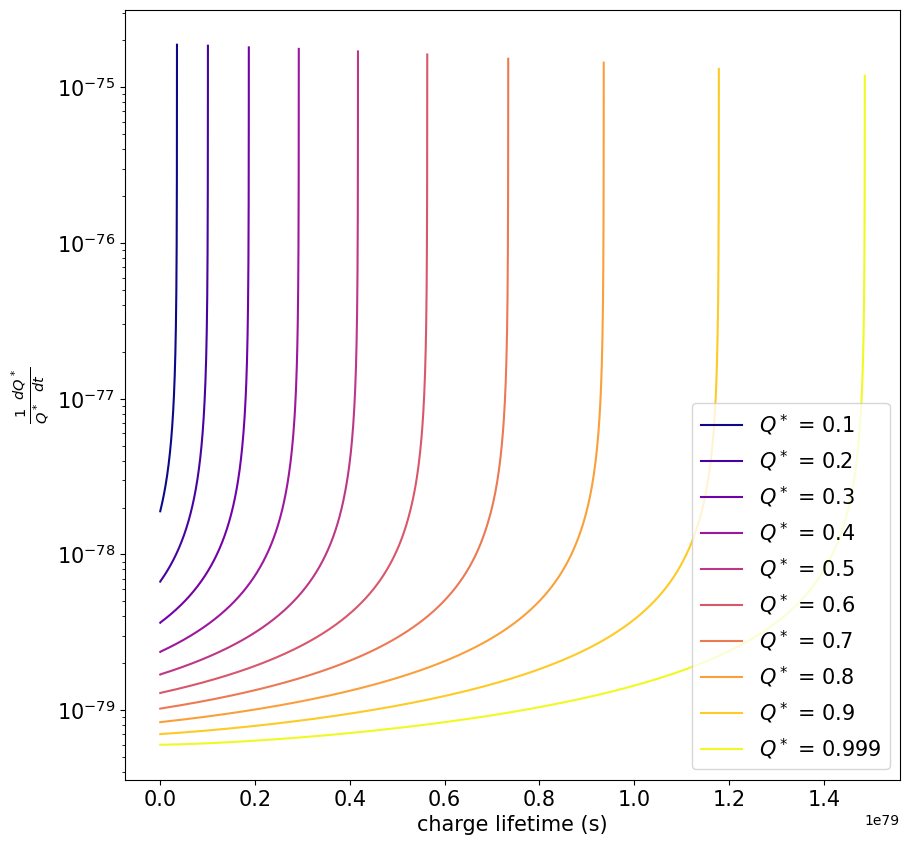}
\caption{Effect of Schwinger pair production on $\frac{dQ^*}{dt}$ for various effective charges for a BH of initial mass $M = 10^4 M_{\odot}$. All charge is lost before any significant amount of mass is lost.}
\label{fig:schwinger}
\end{figure}

\begin{figure}[t!]
\includegraphics[width = 0.4\textwidth, height = 0.4\textwidth ]{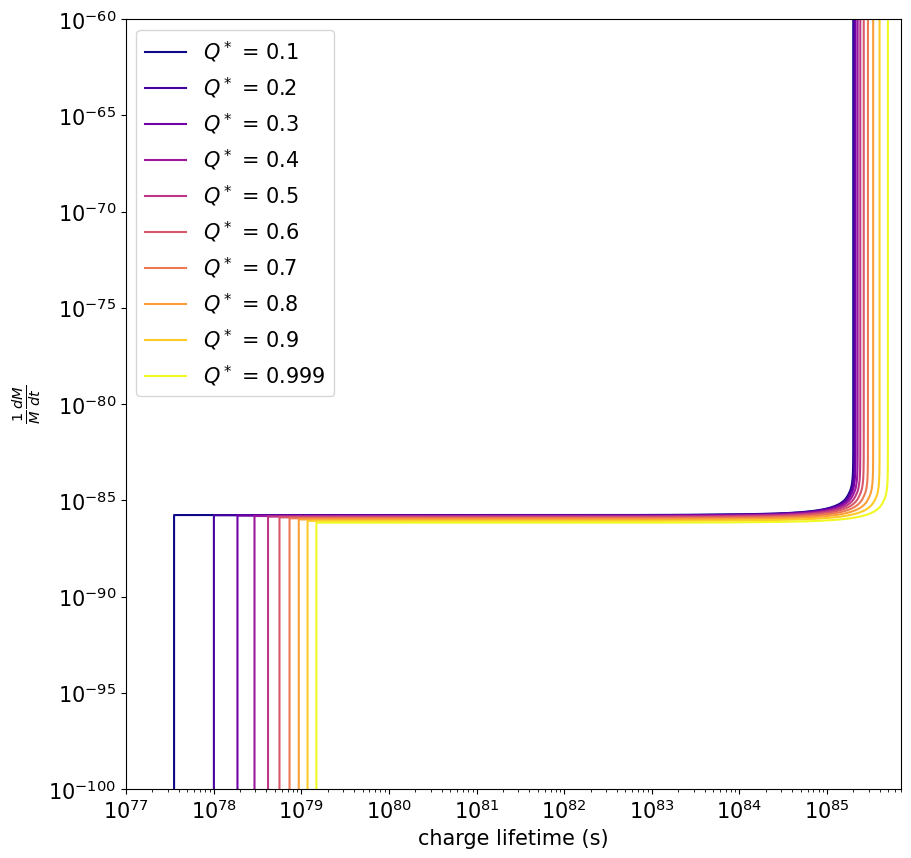}
\caption{Effect of Schwinger pair production of $\frac{dM}{dt}$ for various effective charges. BH has initial mass $M = 10^4 M_{\odot}$. Sudden jump on the left hand side of the plot is when charge is neutralized. From then on BH behaves as if static.}
\label{fig:schwinger_dMdt}
\end{figure}

where $\frac{dQ}{dt}$ describes the charge particle production at the horizon due to quantum vacuum fluctuations and $\frac{dQ_S}{dt}$ describes electron-positron production due to the presence of a large electric field at various $r > r_+$ 
The rapid discharge due to Schwinger pair production would also affect the mass loss rate $\frac{dM}{dt}$ as well. The mass loss rate can be obtained by looking at 
\begin{equation}
    \frac{dM}{dt} = \frac{\partial M}{\partial Q} \frac{\partial Q}{\partial t} = \frac{\partial M}{\partial r_+}\frac{\partial r_+}{\partial Q} \frac{\partial Q}{\partial t}
\end{equation}
Working out the algebra we have the expression 
\begin{equation}
    \frac{dM_S}{dt} = \frac{Q^*M}{r_+}\frac{dQ_S}{dt}
\end{equation}
which is analogous to the result from  \cite{Hiscock:1990ex} in natural units. This additional term will serve to speed up the mass loss rate for a charged BH as long as $\frac{dQ_S}{dt}$ is nonzero. 

We can calculate an order of magnitude estimate to see if this effect will have a significant impact on the discharging of a RN BH. Both the Schwinger and the Hawking discharge rate feature an exponential suppression; in the case of the Schwinger discharge rate this is (Note that this is the discharge rate per unit volume and the hawking rate is per unit energy)
\begin{equation}
    \Gamma_{\rm Sch}\sim \exp\left(-\frac{\pi m_e^2 r_+^2}{e Q}\right),
\end{equation}
with $$r_+=M\left(1+\sqrt{1+(Q/M)^2}\right)$$ and $e$ the electron charge; expressing the above in natural units, neglecting order 1 factors, we get
\begin{equation}
    \Gamma_{\rm Sch}\sim \exp\left(-\frac{m_e^2 M^2}{ e Q}\right).
\end{equation}
The Hawking evaporation rate is Boltzmann suppressed by $m_e/T_H$, where
$$
T_H\sim \frac{1}{M}\left(1-\frac{Q^2}{M^2}\right),
$$

\begin{figure*}[t!]
\begin{center}
    
\includegraphics[width = 0.9\textwidth, height = 0.4\textwidth ]{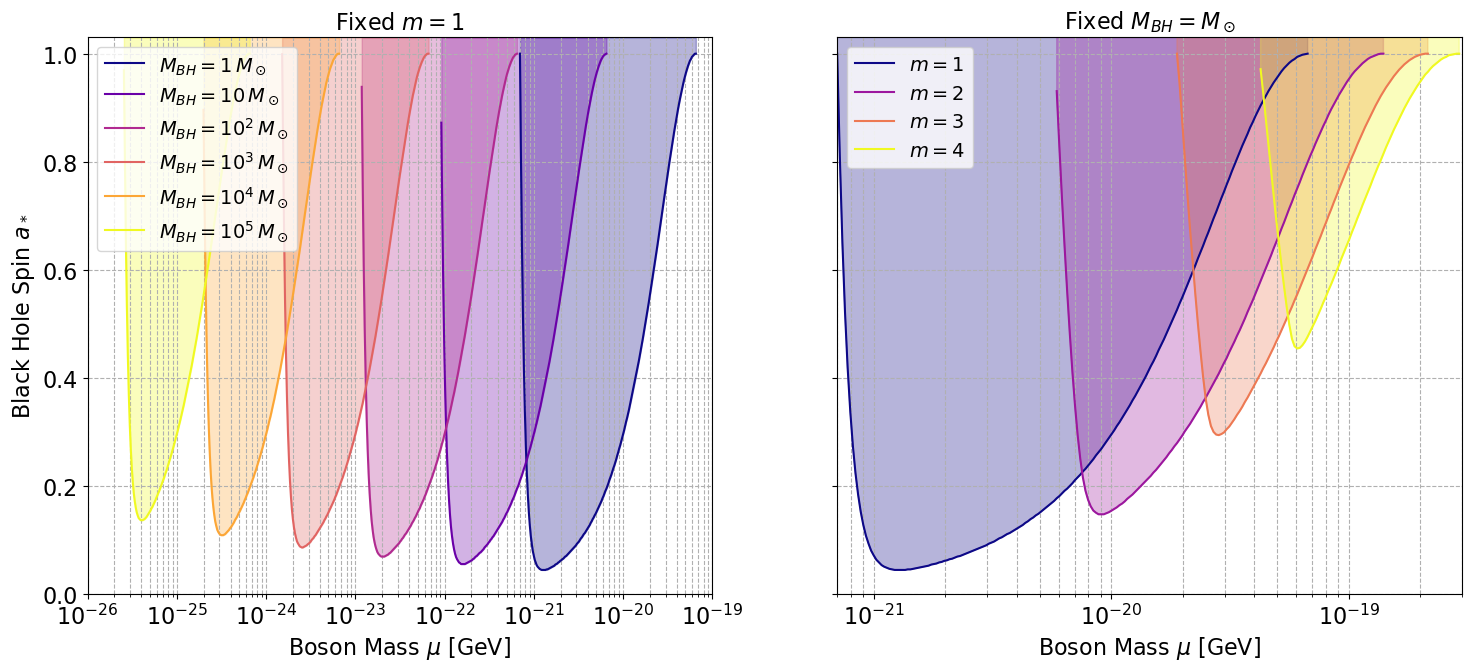}
\caption{ Parameter space for possible superradiance regions for a given black hole mass and light boson.}
\label{fig:SR_Regge}
\end{center}

\end{figure*}

\begin{figure*}[t!]
\begin{center}
    
\includegraphics[width = 0.9\textwidth, height = 0.3\textwidth ]{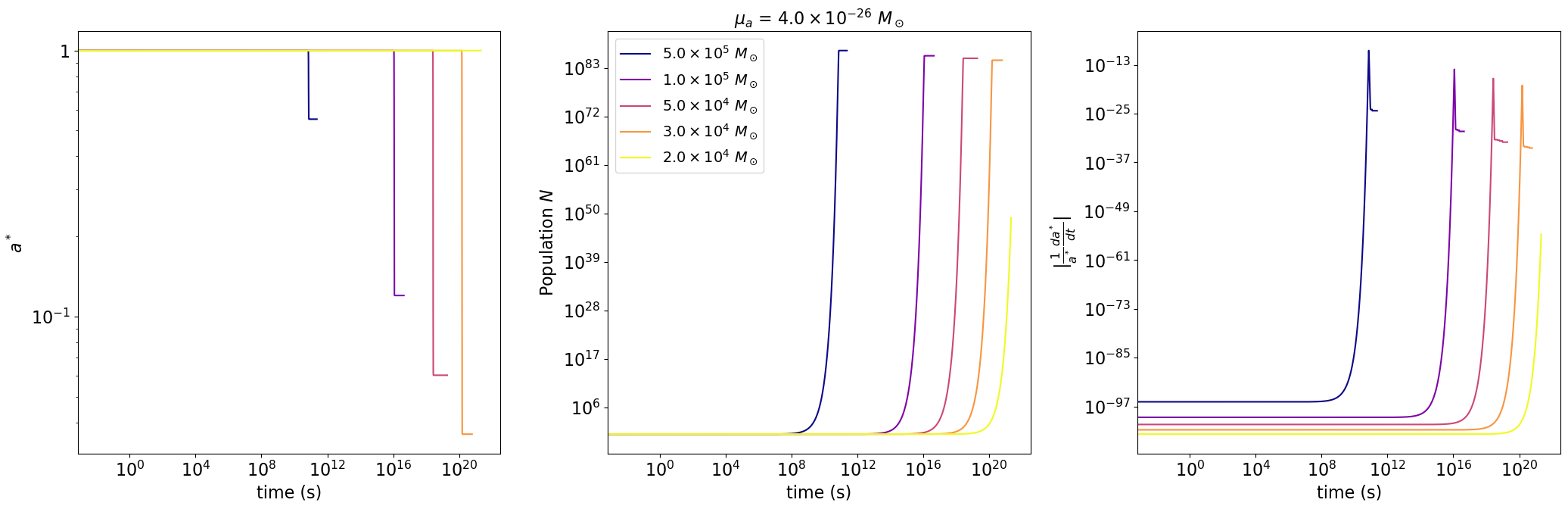}
\caption{Effective spin loss due to superradiance of a Kerr BH. Extracting the majority of spin ($\mathcal{O}(0.1)$ remaining) occurs in approximately $10^{16}$s in comparison to the Hawking lifetime for Kerr BHs of $1.8 \times 10^{24} $ s. }
\label{fig:SR_da}
\end{center}
\end{figure*}

thus
\begin{equation}
    \Gamma_{\rm Hawk}\sim\exp\left(-\frac{M m_e}{1-\frac{Q^2}{M^2}}\right).
\end{equation}
The Schwinger discharge rate dominates when 
\begin{equation}
    \Gamma_{\rm Sch}\gtrsim \Gamma_{\rm Hawk} \Rightarrow \frac{m_e^2 M^2}{ e Q}\lesssim\frac{m_e M}{\left(1-\frac{Q^2}{M^2}\right)}.
\end{equation}
We can rearrange this equation to find that in terms of mass-to-charge ratio $x_e=(e/m_e)$ and $x_{\rm BH}=Q/M$, Schwinger discharge dominates if 
\begin{equation}
    \frac{1-x_{\rm BH}}{x_{\rm BH}}\lesssim x_e\simeq 6.4\times 10^{21}.
\end{equation}
In turn, this implies
\begin{equation}
    x_{\rm BH}\gtrsim 10^{-22}
\end{equation}
Since the Planck charge is 11.7$e$, we find that Schwinger discharge dominates for any black hole with any electric charge $Q\ge e$ when 
$M\le  10^{18}$ grams

Looking at Fig \ref{fig:schwinger}, we can see the how Schwinger pair production enhances the number of charged particles in the Hawking spectra. In this case we have considered a much larger BH ($10^4 M_{\odot}$). The primary purpose of this choice is to consider a stage in a charged BHs life when there is essentially only blackbody radiation. This mechanism only serves to deplete mass from a charged BH and, by Eq \eqref{crossover_eq}, would serve to increase the overall effective charge of such a large BH. Without Schwinger pair production, there would be a tendency for charged BHs to approach extremality in all scenarios. However, we can see that this effect serves to rapidly discharge the BH in comparison to the mass loss timescales. Looking at figure \ref{fig:schwinger_dMdt}, we can see that the lifetime of a BH with various effective charges hovers around $10^{85}$s for BH of this size. This is roughly $10^{16}$ times larger than the timescale for which a BH will discharge due to Schwinger pair production. Its important to note that this analysis only includes effects from the leading order term for Schwinger production and therefore indicates the largest lifetimes possible when including this effect. All higher order terms serve to increase the electron-positron production rate, and therefore will decrease the lifetime of a charged BH. 

\section{Superradiance and Spin: Angular Momentum Extraction }\label{SuperSect} 
Recent work has shown that black hole superradiance can extract angular momentum on timescales much shorter than Hawking emission \cite{tomaselli2024gravitationalatomsblackhole}. Although a black hole cannot be completely spun down through this mechanism, a substantial fraction of its angular momentum can be removed, and we include this effect alongside the Hawking analysis for Kerr black holes.

Superradiance arises from the existence of the ergoregion between the outer and inner horizons, where negative-energy states are permitted \cite{brito2015superradiance}. A particle entering the ergoregion may decay into two modes, one of which carries negative energy $E_1 < 0$. Energy conservation then requires the other mode to emerge with energy $E_2 = E_0 - E_1 > E_0$, so that the outgoing particle carries more energy than the particle originally entering the black hole. The negative-energy mode remains inside the ergoregion and effectively reduces the black hole’s rotational energy.

Superradiance is the amplification of this process: bosonic waves satisfying the superradiant condition grow exponentially by repeatedly extracting rotational energy. The resulting accumulation of a scalar cloud outside the horizon forms a “gravitational atom,” enabling rapid and efficient angular momentum extraction compared to Hawking radiation alone. In the following section we will outline the timescales associated with removing angular momentum from the system via the introduction of massive scalar bosons and the supperadiant effect. 

Superradiance of a BH is allowed so long as the superradiant condition \cite{tomaselli2024gravitationalatomsblackhole}:

\begin{equation}
    \omega < m \Omega_+
    \label{SR_cond}
\end{equation}
is satisfied. In the above equation $\omega$ refers to the frequency of the incoming scalar wave, $m$ the azithmul quantum number of a given state and $\Omega_+$ the angular velocity of the BH horizon. In general, we assume the population of states occurs with $ \omega $ to be roughly the boson mass $ \omega \approx \mu_a$. The overall angular momentum available to be extracted via superradiance is then proportional to the number of particles produced by this process. For each particle extracted from a given BH, the overall angular momentum of the BH is lowered by $\Delta J = m$, or an integer quanta of angular momentum proportional to the given state occupied. Therefore, the overall rate of angular momentum extraction from the BH is equivalent to the rate at which superradiance occurs $\Gamma_{SR}$ for a given particle multiplied by the number of particles of a given state $N_m$:
\begin{equation}
    \frac{dJ}{dt} = -m \Gamma_{SR}N_m
\end{equation}
Again using Eq. \eqref{dadt}, we estimate the change in effective spin as long as we have the mass loss rate due to superradiance. The mass loss rate of the BH is obtained in a similar manner, except we can replace the quanta of spin $m$ by the scalar particle mass $\mu_a$ to get
\begin{equation}
    \frac{dM}{dt} = -\mu_a \Gamma_{SR} N_m.
\end{equation}
All together, the expression for effective spin loss due to superradiance is given by;
\begin{equation}
    \frac{da^*_{SR}}{dt} =  -\frac{1}{M^2} m\Gamma_{SR}N_m + 2\frac{a^*}{M}   \mu_a \Gamma_{SR}N_m.
\end{equation}
This equation is expected to be valid so long as the superradiance condition in Eq. \eqref{SR_cond} is satisfied and the superradiance rate $\Gamma_{SR}$ is faster than the Eddinton accretion time. We have simulated the effective spin available to be extracted from the BH in the Regge trajectory plots in Fig. \ref{fig:SR_Regge}. For any BH mass we can see that, for quantum numbers $\{n,l,m\}$, the $m=1$ state allows for the largest $\Delta a^*$ in the right hand panel. Additionally, we simulate the parameter space for various BH and scalar boson masses that allow for superradiance. We can see as a general trend that as BH mass increases the overall spin available for extraction decreases. Additionally, we expect that constraints on light boson candidates to be another limiting factor in the possibility of  this SR regime being reached for heavy BHs. 
\begin{figure}[t!]
\begin{center}
    
\includegraphics[width = 0.4\textwidth, height = 0.4\textwidth ]{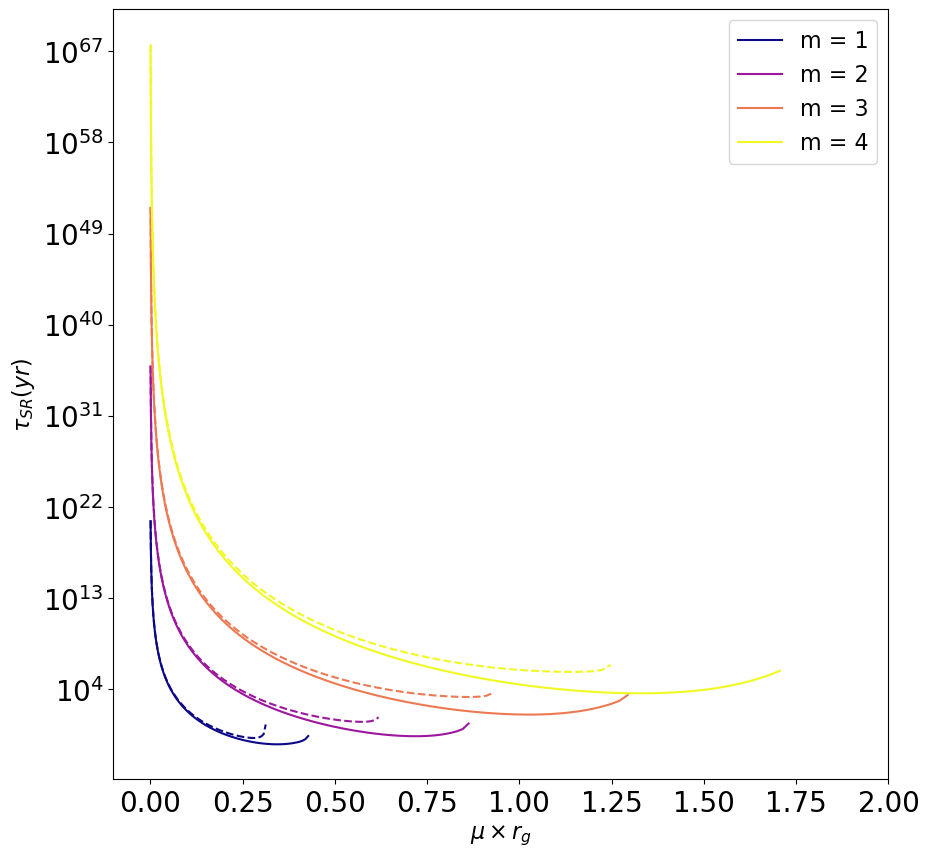}
\caption{Superradiant event lifetime for a given BH and boson mass, set by the gravitational fine structure constant $\alpha = \mu_a \times r_g$. We use $M = 10^5 M_\odot$. Solid lines correspond to $a^* = 0.99$ and dotted to $a^* = 0.9$. Lifetimes are expected to increase proportionally as mass increases, as there will be more angular momentum available for a given effective spin.  }
\label{fig:SR_life}
\end{center}

\end{figure}

Still, we can estimate the spin down lifetime of a BH due to superradiance with Eq. \ref{SR_cond}. Utilizing the results of \cite{tomaselli2024gravitationalatomsblackhole}, the $\{n,l,m\} = \{2,1,1\}$ state is the most efficient at extracting angular momentum.  In general the lifetime of a BHs superradiant event of a BH can be estimated as $\tau_{SR} = \frac{1}{\Gamma_{SR}}$ \cite{Arvanitaki_2015}. Using our superradiant rates derived from \cite{tomaselli2024gravitationalatomsblackhole}, we illustrate the expected lifetime of a given superradiant event for a solar mass BH and variable gravitational fine structure constant $\alpha =  \mu_a M$ in Fig. \ref{fig:SR_life}. The shortest spin-down time we find is approximately $10^{16}\,\mathrm{s}$, corresponding to a reduction from $a^* = 0.9999$ to $a^* \approx 0.15$ for a $10^5 M_\odot$ black hole. This is approximately $10^8$ times faster than it takes for a Kerr BH to lose spin due to purely radiative processes. Granted, this is still a rotating BH, but this process serves to greatly reduce the spin and overall lifetime due to the significant influence of effective spin on BH temperature and particle emission rate. For the $m = 1$ superradiant states, smaller values of the gravitational fine-structure constant $\alpha$ correspond to more spin being extractable. This explains why the superradiant lifetime grows as $\alpha = \mu_a r_g$ approaches zero in the figure Overall, a significant amount of the available spin can be lost from a superradiant event. When considering the lowest superradiant state, all BHs shed spin under approximately $10^{20}$ seconds, which is significantly faster than the expected Kerr evaporation timescales.

\section{Conclusions}\label{sec:conclusions}

This comprehensive study systematically compares the evaporation dynamics of Schwarzschild, Kerr, and Reissner-Nordström  black holes through the lens of Page factors, revealing critical insights into the interplay of mass, charge, and angular momentum loss. For {\bf Schwarzschild black holes}, isotropic Hawking radiation drives mass loss on timescales governed solely by the Stefan-Boltzmann law, with evaporation rates scaling as $ M^{-2} $. In contrast, {\bf Kerr black holes} exhibit rapid angular momentum dissipation due to preferential emission of high-spin particles and superradiance effects, leading to a spin-down phase orders of magnitude faster than mass loss. This asymmetry arises from frame-dragging effects, which amplify emission from equatorial regions and enhance the coupling of angular momentum to radiation fields.  

For {\bf RN black holes}, charge loss dominates early-stage evaporation, mediated by the efficient emission of charged particles through both Hawking radiation and Schwinger pair production. The Coulomb potential barrier introduces a self-regulating mechanism: high charge-to-mass ratios ($ Q_*/M $) suppress thermal emission by lowering the black hole temperature, while simultaneously accelerating charge loss via enhanced particle tunneling. Near-extremal configurations ($ Q_* \rightarrow 1 $) exhibit prolonged lifetimes due to suppressed Hawking temperatures, though quantum corrections destabilize these states over cosmological timescales.  

A pivotal finding is the {\bf modification of evaporation hierarchies} by non-Standard Model  particle species. Introducing large numbers of uncharged dark matter degrees of freedom shifts the balance between mass and charge loss rates. For example, a dark sector with $ \mathcal{O}(100) $ massless particles can invert the canonical hierarchy, causing $ \dot{M}/M $ to exceed $ \dot{Q}_*/Q_* $ even for highly charged black holes. This challenges assumptions about "charge neutrality" as a universal endpoint of evaporation and suggests that astrophysical black holes in dark matter-rich environments may retain residual charges longer than predicted by SM-only models.  

The role of {\bf greybody factors} further complicates these dynamics. For Kerr black holes, frame-dragging distorts emission spectra, favoring low-angular-momentum modes and accelerating spin loss. In RN systems, greybody corrections suppress charged particle emission at low energies but enhance high-energy discharges, creating a feedback loop that regulates charge loss. These effects are particularly pronounced in near-extremal regimes, where geometric optics approximations break down, necessitating full wave-equation treatments for accuracy.  

{\bf Schwinger pair production} emerges as a critical mechanism for light ($ M \lesssim 10^{17} \, \mathrm{g} $), highly charged black holes. At $ Q_*/M \gtrsim 0.5 $, this quantum electrodynamic process dominates over Hawking radiation, ejecting $ e^\pm $ pairs and accelerating charge loss. However, for primordial black holes formed with near-extremal charges, Schwinger emission can transiently *increase* $ Q_* $ by preferentially removing opposite charges—a counterintuitive result with implications for early-universe baryogenesis.  

{\bf Superradiance} plays a role for any mass BH to extract significant amounts of angular momentum of a Kerr BH in the form of a light boson cloud. For a given BH mass, a range of axion masses allow for $\Delta a^* = \mathcal{O}(1)$ changes in the overall angular momentum on timescales much faster than that of Hawking radiation. The possibility of forming a gravitational atom allows for much more rapid increases in temperature of the BH and speeds up the overall mass loss of the system as well.

These results underscore the inadequacy of single-parameter evaporation models and highlight the need for {\bf multivariate frameworks} incorporating particle physics beyond the SM. Future work should explore the interplay between evaporation and accretion in astrophysical environments, quantum gravity corrections to semi-classical approximations, and observational signatures of charged or spinning black holes in gravitational wave or cosmic ray data. Additionally, the discovery that dark sectors can dramatically alter evaporation hierarchies invites experimental tests via high-precision measurements of black hole merger remnants or gamma-ray burst spectra. By bridging thermodynamics, quantum field theory, and astrophysics, this study advances our understanding of black holes as multichannel dissipative systems and provides tools to probe fundamental physics through their evaporation.

\newpage

\bibliographystyle{unsrt} 
\bibliography{bibliography}

@article{hawking1975particle,
  author = {Hawking, S. W.},
  title = {Particle creation by black holes},
  journal = {Communications in Mathematical Physics},
  volume = {43},
  number = {3},
  pages = {199-220},
  year = {1975}
}

@article{page1976massless,
  author = {Page, Don N.},
  title = {Particle emission rates from a black hole. I. Massless particles from uncharged black holes},
  journal = {Physical Review D},
  volume = {13},
  number = {2},
  pages = {198-206},
  year = {1976}
}

@article{page1976rotating,
  author = {Page, Don N.},
  title = {Particle emission rates from a black hole. II. Massless particles from a rotating hole},
  journal = {Physical Review D},
  volume = {14},
  number = {12},
  pages = {3260-3273},
  year = {1976}
}

@article{page1977charged,
  author = {Page, Don N.},
  title = {Particle emission rates from a black hole. III. Charged leptons from a nonrotating hole},
  journal = {Physical Review D},
  volume = {16},
  number = {8},
  pages = {2402-2411},
  year = {1977}
}

@article{taylor1998evaporation,
  author = {Taylor, B. and Hiscock, W.},
  title = {Evaporation of a Kerr black hole by emission of scalar and higher spin particles},
  journal = {Physical Review D},
  volume = {55},
  pages = {6116-6125},
  year = {1998}
}

@article{masina2021dark,
  author = {Masina, Isabella},
  title = {Dark matter and dark radiation from evaporating Kerr primordial black holes},
  journal = {Gravitation and Cosmology},
  volume = {27},
  pages = {134-147},
  year = {2021}
}

@article{dong2015gravitational,
  author = {Dong, R. and Stojkovic, D.},
  title = {Gravitational wave production by Hawking radiation from rotating primordial black holes},
  journal = {Journal of Cosmology and Astroparticle Physics},
  year = {2015}
}

@article{boonserm2012bounding,
  author = {Ngampitipan, T. and Boonserm, P.},
  title = {Bounding the greybody factors for non-rotating black holes},
  journal = {Classical and Quantum Gravity},
  volume = {29},
  year = {2012}
}

@article{hod2019penrose,
  author = {Hod, Shahar},
  title = {Hawking radiation may violate the Penrose cosmic censorship conjecture},
  journal = {International Journal of Modern Physics D},
  year = {2019}
}

@article{brown2024evaporation,
  author = {Brown, A. and Usatyuk, M.},
  title = {The evaporation of charged black holes},
  journal = {Not Provided},
  year = {2024}
}

@article{hawking1997extremal,
  author = {Hawking, S. W. and Taylor-Robinson, M.},
  title = {Evolution of near extremal black holes},
  journal = {Physical Review D},
  year = {1997}
}

@article{Blackhawk,
   title={BlackHawk: a public code for calculating the Hawking evaporation spectra of any black hole distribution},
   volume={79},
   ISSN={1434-6052},
   url={http://dx.doi.org/10.1140/epjc/s10052-019-7161-1},
   DOI={10.1140/epjc/s10052-019-7161-1},
   number={8},
   journal={The European Physical Journal C},
   publisher={Springer Science and Business Media LLC},
   author={Arbey, Alexandre and Auffinger, Jérémy},
   year={2019},
   month=aug }

@article{Page_rotating,
  title = {Particle emission rates from a black hole. II. Massless particles from a rotating hole},
  author = {Page, Don N.},
  journal = {Phys. Rev. D},
  volume = {14},
  issue = {12},
  pages = {3260--3273},
  numpages = {0},
  year = {1976},
  month = {Dec},
  publisher = {American Physical Society},
  doi = {10.1103/PhysRevD.14.3260},
  url = {https://link.aps.org/doi/10.1103/PhysRevD.14.3260}
}

@article{Hawking_BHE,
	author = {HAWKING, S.  W. },
	journal = {Nature},
	number = {5443},
	pages = {30--31},
	title = {Black hole explosions?},
	volume = {248},
	year = {1974}}

@article{Hiscock:1990ex,
  author         = "Hiscock, W. A. and Weems, L. D.",
  title          = "{Evolution of Charged Black Holes}",
  journal        = "Phys. Rev. D",
  volume         = "41",
  year           = "1990",
  pages          = "1142",
  doi            = "10.1103/PhysRevD.41.1142",
}

@article{Chen:2012zn,
    author = "Chen, Chiang-Mei and Wu, Ming-Fan",
    title = "{Spontaneous Pair Production in Reissner-Nordström Black Holes}",
    journal = "Phys. Rev. D",
    volume = "85",
    year = "2012",
    pages = "124041",
    doi = "10.1103/PhysRevD.85.124041",
    citations = "51"
}

@article{Lin:2024aa,
    author = "Lin, Puxin and Shiu, Gary",
    title = "{Schwinger Effect of Extremal Reissner-Nordström Black Holes}",
    year = "2024",
    journal = "Not Provided",
    citations = "1"
}

@article{Schwinger_seminal,
  title = {On Gauge Invariance and Vacuum Polarization},
  author = {Schwinger, Julian},
  journal = {Phys. Rev.},
  volume = {82},
  issue = {5},
  pages = {664--679},
  numpages = {0},
  year = {1951},
  month = {Jun},
  publisher = {American Physical Society},
  doi = {10.1103/PhysRev.82.664},
  url = {https://link.aps.org/doi/10.1103/PhysRev.82.664}
}

@misc{tomaselli2024gravitationalatomsblackhole,
      title={Gravitational Atoms and Black Hole Binaries}, 
      author={Giovanni Maria Tomaselli},
      year={2024},
      eprint={2412.12526},
      archivePrefix={arXiv},
      primaryClass={gr-qc},
      url={https://arxiv.org/abs/2412.12526}, 
}

@book{brito2015superradiance,
  title={Superradiance: Energy Extraction, Black-Hole Bombs and Implications for Astrophysics and Particle Physics},
  author={Brito, R. and Cardoso, V. and Pani, P.},
  isbn={9783319190006},
  series={Lecture Notes in Physics},
  url={https://books.google.com/books?id=gGonCgAAQBAJ},
  year={2015},
  publisher={Springer International Publishing}
}

@article{Arvanitaki_2015,
   title={Discovering the QCD axion with black holes and gravitational waves},
   volume={91},
   ISSN={1550-2368},
   url={http://dx.doi.org/10.1103/PhysRevD.91.084011},
   DOI={10.1103/physrevd.91.084011},
   number={8},
   journal={Physical Review D},
   publisher={American Physical Society (APS)},
   author={Arvanitaki, Asimina and Baryakhtar, Masha and Huang, Xinlu},
   year={2015},
   month=apr }

\end{document}